\documentclass[aps,prd,showpacs,preprint,groupedaddress]{revtex4}

\begin{document}
\title{Particle Radiation From Gibbons-Maeda Black Hole }
\author{Heng-Zhong FANG}
\email{qdfang@163.com}
\affiliation{Department of Physics, Ocean
University of China, Qingdao 266071, China}
\author{Jian-Yang Zhu}
\affiliation{Department of Physics, Beijing Normal University,
Beijing 100875,China}
\date{\today }

\begin{abstract}
This paper investigates the particle radiation from Gibbons-Maeda black
hole. Taking into account the self-gravitation of the particle, we calculate
the tunnelling rate of the massless particle across the horizon, then we
promote the work to the radiation of the charged particle. The calculations
prove that the rate of tunnelling equals precisely the exponent of the
difference of the black hole entropy before and after emission and the
radiation spectrum deviates from exact thermal. The conclusion supports the
viewpoint of information conservation.
\end{abstract}

\pacs{04.70.Dy}

\maketitle
\section{Introduction}

For a long time, people believed that the black hole was such an object that
things approaching it should be absorbed, while anything inside it could not
escape, the black hole could only grow bigger and bigger. However, in
1970's, S.W.Hawking declared\cite{Hawk} his astonishing discovery that the
black hole can radiate particles. Hawking's idea is that when a virtual
particle pair are created near the horizon, the negative energy particle is
absorbed by the black hole while the positive tunnels out of the horizon,
then it materializes as a real one, appearing as Hawking radiation. The
black hole radiation is purely of a quantum effect, the emission is a
procedure of particle tunnelling across the potential barrier.\ Gibbons and
Hawking also demonstrated\cite{Gibb} that the energy spectrum of the
particle radiation from a collapsing star is of exactly thermal. From then
on, so many studies about the Hawking radiation from various black holes
gave the same conclusion that the radiation spectrum is of black body
spectrum. But there are two puzzles. One is the barrier, where the barrier
exists is not clear. The other is the radiation spectrum, because we can get
only one parameter from a thermal spectrum, the temperature. So if the
energy spectrum is exactly thermal, the radiation should not bring us any
macroscopic information about the material in the black hole, even that once
the black hole has evaporated out, there would be no any marks left, all the
information including the unitary would be lost. In fact, the argument about
the information loss has lasted many years.

M.K.Parikh and F.Wilczek\cite{MK2,MK3} claimed that there is no a barrier
existing before the particle tunnels out of the black hole, instead, the
barrier is created by the outgoing particle itself. The total energy of a
stationary space-time is conservational during the emission, this demands
that, when a particle is radiated, the left mass of the black hole decrease,
so the radius of the horizon contracts from its original radius to a new,
smaller radius, and the scale of the contraction depends on the energy of
the radiated particle, the separation between the initial and final radius
is just the barrier for tunnelling. During the course of the radiation, if
the self-gravitation of the particle is taken into account , the space-time
geometry is dynamic, the black hole can be regarded as an excited metastable
state. M.K.Parikh and F.Wilczek presented a derivation of massless particle
radiation and obtained that the emission spectrum of schwarzschild black
hole deviates from the exact thermal. Enlightened by M.K.Parikh, many works%
\cite{vag,fang,zhang1,zhang2,fang1} were performed about the particle radiation
from different black holes. In \cite{zhang1,zhang2}, the radiation of
massive and charged particle were investigated, to get the velocity of a
massive particle, the authors used the Laudau's condition of the clock
synchronization, for the radiation of charged particle, they regarded the
electromagnetic potential as an ignorable coordinate and considered the
matter-gravity energy and electro-magnetic field energy separately.

In this paper, we promote the work of Parikh and Wilczek to the radiation of
Gibbons-Maeda Black hole, calculate the spectrum of particle radiation in a
completely different way. The four-velocity normalization equation $%
U^aU_a=-1 $ are utilized to get the velocity of the massive particle. The
total energy of the space-time is considered. Both of the calculations for
the radiation of massless particle and charged particles give the same
conclusion that the rate of tunnelling equals precisely the exponent of the
difference of the entropy of the black hole before and after emission, this
implies that the information is preserved during the black hole radiation.

\section{A convenient coordinate system for GM black hole}

The line element of Gibbons-Maeda(GM) black hole is described\cite{GM} by
\begin{equation}
ds^2=-%
%TCIMACRO{\dfrac{(r-r_{+})(r-r_{-})}{R^2} }
%BeginExpansion
{\displaystyle {(r-r_{+})(r-r_{-}) \over R^2}}%
%EndExpansion
dt^2+%
%TCIMACRO{\dfrac{R^2}{(r-r_{+})(r-r_{-})} }
%BeginExpansion
{\displaystyle {R^2 \over (r-r_{+})(r-r_{-})}}%
%EndExpansion
dr^2+R^2(d\theta ^2+\sin ^2\theta d\varphi ^2)\text{,}  \label{c1}
\end{equation}
where $R^2=r^2-D^2$, $r_{+}$ and $r_{-}$ are respectively the event horizon
and the inner horizon, satisfy
\begin{equation}
r_{\pm }=M\pm \sqrt{M^2+D^2-P^2-Q^2}\text{,}  \label{c2}
\end{equation}
and the axion dilaton charge $D$ is
\begin{equation}
D=%
%TCIMACRO{\dfrac{P^2-Q^2}{2M} }
%BeginExpansion
{\displaystyle {P^2-Q^2 \over 2M}}%
%EndExpansion
\text{,}  \label{c3}
\end{equation}
the parameters $P$ and $Q$ represent the black hole magnetic charge and the
electric charge, respectively.

It is easy to find that one of the components of the metric in (\ref{c1}) is
singular at $r=r_{+}$. In order to describe the particle tunnelling across
the horizon, we need a coordinate in which the components of the metric and
the inverse metric do not diverge at both sides of the horizon. Perform such
a coordinate transformation that let $dT=dt+f(r)dr$, then the line element
of GM black hole is written as
\begin{equation}
ds^2=-(1-g)\Delta dT^2+2(1-g)\Delta f(r)dTdr+\left[ \frac 1{(1-g)\Delta }%
-(1-g)\Delta f^2(r)\right] dr^2+R^2d\Omega ^2\text{,}  \label{c4}
\end{equation}
where $g={\displaystyle {r_{+} \over r}}$, and $\Delta =
{\displaystyle {r^2(1-{\displaystyle {r_{-} \over r}}) \over R^2}}$.
So in coordinate system $(T,r,\theta ,\varphi )$,
\begin{equation}
g_{rr}=\frac 1{(1-g)\Delta }-(1-g)\Delta f^2(r)=\frac{1-(1-g)^2\Delta
^2f^2(r)}{(1-g)\Delta }\text{.}  \label{c5}
\end{equation}
In order to eliminate the singularity of $g_{rr}$ at $r=r_{+}$, taking into
account the dimension, we suppose that
\begin{equation}
1-(1-g)^2\Delta ^2f^2(r)=k(1-g)\text{,}  \label{c6}
\end{equation}
then we have
\begin{equation}
f(r)=\frac{\sqrt{(1-k)+kg}}{(1-g)\Delta }\text{,}  \label{c7}
\end{equation}
for the simplicity, let $k=1$, get
\begin{equation}
f(r)=\frac{\sqrt{g}}{(1-g)\Delta }\text{.}  \label{c8}
\end{equation}
Substitute (\ref{c8}) to (\ref{c4}), we get that
\begin{equation}
ds^2=-(1-g)\Delta dT^2+2\sqrt{g}dTdr+\frac 1\Delta dr^2+R^2d\Omega ^2\text{.}
\label{c9}
\end{equation}

From (\ref{c9}) we find that the new coordinate system $(T,r,\theta ,\varphi
)$ has many attractive features. First, each component of the metric and the
inverse metric is well behaved at the horizon $r=r_{+}$, it is important for
the calculation of the particle tunnelling across the horizon. Second, in
the global space-time there exists a Killing vector ($\frac \partial {%
\partial T})^a$, so we can construct the constant-time slices which are
Euclidean flat space.

\section{The radiation of massless particles}

From equation (\ref{c9}), a massless particle moves along the radial null
geodesic obeys
\begin{equation}
\stackrel{.}{r}=\frac{dr}{dT}=-\Delta \sqrt{g}\pm \Delta =%
%TCIMACRO{\dfrac{r^2}{R^2} }
%BeginExpansion
{\displaystyle {r^2 \over R^2}}%
%EndExpansion
(1-%
%TCIMACRO{\dfrac{r_{-}}r }
%BeginExpansion
{\displaystyle {r_{-} \over r}}%
%EndExpansion
)(\pm 1-\sqrt{\frac{r_{+}}r})\text{,}  \label{c10}
\end{equation}
where the upper(lower) sign corresponds the outgoing(ingoing) geodesic under
the implicit assumption that $T$ increase towards the future, here we take
the upper sign. Because the metric is spherical symmetry, so regarding the
outgoing particle as a s-wave i.e. a shell of energy is acceptable. For GM
space-time, the ADM mass is just the parameter mass, $M$, it is
conservational. When a shell of energy $\omega $ tunnels out of the horizon,
the leftover mass of the black hole becomes $M-\omega $. To the radiated
particle, the geometry of the space-time is changed, so the equation (\ref
{c10}) should be modified by replacing $r_{\pm }(M)$ with $r_{\pm }(M-\omega
)$.

Suppose that the outgoing wave were traced back towards the horizon, its
wave-length, measured by local observers, would be blue-shifted, near the
horizon, the radial wave number approaches infinity, so that the geometrical
optics limit, or WKB approximation is valid. In the semi-classical limit,
according to the WKB approximation, the s-wave of the outgoing positive
energy particle can be expressed as $\Psi (r)=e^{iI(r)}$, and the
probability of the tunnelling $\Gamma $ takes the form
\begin{equation}
\Gamma \sim \left| \Psi (r)\right| ^2=\exp \left( -2%
%TCIMACRO{\func{Im} }
%BeginExpansion
\mathop{\rm Im}%
%EndExpansion
I\right) \text{,}  \label{c11}
\end{equation}
here $I$ is the action, it can be obtained from

\begin{equation}
I=\int Ldt=\int p_rdr-\int Hdt,  \label{c12}
\end{equation}
where $L$ and $H$ are respectively the lagrangian and the Hamiltonian of the
particle, $p_r$ is the radial momentum. Because $H$ is real, from (\ref{c11}%
) we know that it has no contribution to $\Gamma $, so the second term of
the right hand side of equation (\ref{c12}) is ignored in the following
discussion, that
\begin{equation}
I=\int_{r_i}^{r_f}p_rdr=\int_{r_i}^{r_f}\int_0^{p_r}dp_r^{^{\prime }}dr\text{%
,}  \label{c13}
\end{equation}
where the initial radius $r_i$ correspondences the site of pair-creation,
which should be slightly inside the initial horizon, $r_i\approx r_{+}(M_i)$%
, while the final radius $r_f$, to be slightly outside the final position of
the horizon, $r_f\approx r_{+}(M_f)$, $M_i\ $and $M_f$ are respectively the
initial and final mass of the black hole, $r_{f\text{ }}$is actually less
than $r_i$. The separation between $r_{i\text{ }}$and $r_f$ forms the
barrier. Changing variable from momentum to energy by using Hamilton's
equation $\frac{dH}{dp}\mid _r=\frac{\partial H}{\partial p}=\stackrel{.}{r}$%
, we get
\begin{equation}
I=\int_{r_i}^{r_f}\int_0^H%
%TCIMACRO{\dfrac{dH^{\prime }}{\dot{r}} }
%BeginExpansion
{\displaystyle {dH^{\prime } \over \dot{r}}}%
%EndExpansion
dr=\int_{r_i}^{r_f}\int_{M_i}^{M_f}%
%TCIMACRO{\dfrac{dM}{\dot{r}} }
%BeginExpansion
{\displaystyle {dM \over \dot{r}}}%
%EndExpansion
dr=\int_{M_i}^{M_f}\int_{r_i}^{r_f}%
%TCIMACRO{\dfrac{R^2dMdr}{r^2(1-\dfrac{r_{-}}r)(1-\sqrt{\dfrac{r_{+}}r})} }
%BeginExpansion
{\displaystyle {R^2dMdr \over r^2(1-{\displaystyle {r_{-} \over r}})(1-\sqrt{{\displaystyle {r_{+} \over r}}})}}%
%EndExpansion
\text{.}  \label{c14}
\end{equation}
Substituting $u=\sqrt{r}$ to (\ref{c14}), we get
\begin{equation}
I=\int_{M_i}^{M_f}\int_{\sqrt{r_i}}^{\sqrt{r_f}}%
%TCIMACRO{\dfrac{2(u^4-D^2)dMdu}{(u^2-r_{-})(u-\sqrt{r_{+}})} }
%BeginExpansion
{\displaystyle {2(u^4-D^2)dMdu \over (u^2-r_{-})(u-\sqrt{r_{+}})}}%
%EndExpansion
\text{.}  \label{c15}
\end{equation}
It is easy to find that the integrand has a pole at $u=\sqrt{r_{+}}$. In
order to integrate across the singularity $r=r_{+}$, replacing $\sqrt{r_{+}}$
by $\sqrt{r_{+}}-i\epsilon $, we have
\begin{equation}
I=\lim_{\epsilon \rightarrow 0}\int_{M_i}^{M_f}\int_{\sqrt{r_i}}^{\sqrt{r_f}}%
%TCIMACRO{\dfrac{2(u^4-D^2)dMdu}{(u^2-r_{-})(u-\sqrt{r_{+}}+i\epsilon )} }
%BeginExpansion
{\displaystyle {2(u^4-D^2)dMdu \over (u^2-r_{-})(u-\sqrt{r_{+}}+i\epsilon )}}%
%EndExpansion
=\lim_{\epsilon \rightarrow 0}\int_{M_i}^{M_f}\int_{\sqrt{r_i}}^{\sqrt{r_f}}%
%TCIMACRO{\dfrac{f(u)dMdu}{u-\sqrt{r_{+}}+i\epsilon } }
%BeginExpansion
{\displaystyle {f(u)dMdu \over u-\sqrt{r_{+}}+i\epsilon }}%
%EndExpansion
\text{,}  \label{c16}
\end{equation}
where $f(u)=%
%TCIMACRO{\dfrac{2(u^4-D^2)}{(u^2-r_{-})}}
%BeginExpansion
{\displaystyle {2(u^4-D^2) \over (u^2-r_{-})}}%
%EndExpansion
$. From (\ref{c11}), we know that the real part of $I$ only contributes a
phase while the imaginary part contributes the amplitude of the tunnelling
rate, so what we are interested in is only the imaginary part of $I$.
According to the Feynman prescription, the integral can be evaluated by
deforming the contour around the pole $\sqrt{r_{+}}-i\epsilon $. So we get
\begin{equation}
%TCIMACRO{\func{Im} }
%BeginExpansion
\mathop{\rm Im}%
%EndExpansion
I=-\pi \int_{M_i}^{M_f}f(\sqrt{r_{+}})dM=-\pi \int_{M_i}^{M_f}%
%TCIMACRO{\dfrac{2(r_{+}^2-D^2)}{(r_{+}-r_{-})} }
%BeginExpansion
{\displaystyle {2(r_{+}^2-D^2) \over (r_{+}-r_{-})}}%
%EndExpansion
dM\text{.}  \label{c17}
\end{equation}
From (\ref{c2}), we have
\begin{equation}
%TCIMACRO{\func{Im} }
%BeginExpansion
\mathop{\rm Im}%
%EndExpansion
I=-\pi \int_{M_i}^{M_f}%
%TCIMACRO{\dfrac{(M^2-MD-Q^2+M\sqrt{(M-D)^2-2Q^2})}{\sqrt{(M-D)^2-2Q^2}} }
%BeginExpansion
{\displaystyle {(M^2-MD-Q^2+M\sqrt{(M-D)^2-2Q^2}) \over \sqrt{(M-D)^2-2Q^2}}}%
%EndExpansion
dM=-\int_{M_i}^{M_f}K(M)dM\text{,}  \label{c18}
\end{equation}
where
\begin{equation}
K(M)=\pi
%TCIMACRO{
%\dfrac{\left( M^2-MD-Q^2+M\sqrt{(M-D)^2-2Q^2}\right) }{\sqrt{(M-D)^2-2Q^2}} }
%BeginExpansion
{\displaystyle {\left( M^2-MD-Q^2+M\sqrt{(M-D)^2-2Q^2}\right)  \over \sqrt{(M-D)^2-2Q^2}}}%
%EndExpansion
\text{.}  \label{c19}
\end{equation}

For Gibbons-Maeda black hole, the Bekenstein-Hawking entropy $S$ is\cite{GAO}
\begin{equation}
S=\pi (r_{+}^2-D^2)\text{.}  \label{c20}
\end{equation}
So
\begin{eqnarray}
%TCIMACRO{\dfrac{\partial S}{\partial M} }
%BeginExpansion
{\displaystyle {\partial S \over \partial M}}%
%EndExpansion
&=&\pi \left[ 2r_{+}%
%TCIMACRO{\dfrac{\partial r_{+}}{\partial M}}
%BeginExpansion
{\displaystyle {\partial r_{+} \over \partial M}}%
%EndExpansion
-2D%
%TCIMACRO{\dfrac{\partial D}{\partial M}}
%BeginExpansion
{\displaystyle {\partial D \over \partial M}}%
%EndExpansion
\right]   \nonumber \\
&=&%
%TCIMACRO{\dfrac{2\pi }{\sqrt{M^2+D^2-P^2-Q^2}}}
%BeginExpansion
{\displaystyle {2\pi  \over \sqrt{M^2+D^2-P^2-Q^2}}}%
%EndExpansion
\left( M^2-MD-Q^2+M\sqrt{(M-D)^2-2Q^2}\right)   \label{c21}
\end{eqnarray}
Comparing (\ref{c19}) with (\ref{c21}), we are surprised to find that $%
%TCIMACRO{\dfrac{\partial S}{\partial M}}
%BeginExpansion
{\displaystyle {\partial S \over \partial M}}%
%EndExpansion
=2K(M).$ So that
\begin{equation}
%TCIMACRO{\func{Im}}
%BeginExpansion
\mathop{\rm Im}%
%EndExpansion
I=-\frac 12\int_{M_i}^{M_f}%
%TCIMACRO{\dfrac{\partial S}{\partial M}}
%BeginExpansion
{\displaystyle {\partial S \over \partial M}}%
%EndExpansion
dM=-\frac 12\int_{S_i}^{S_f}dS=-\frac 12\Delta S  \label{c22}
\end{equation}
Then the rate of emission satisfies
\begin{equation}
\Gamma \sim \exp (-2%
%TCIMACRO{\func{Im}}
%BeginExpansion
\mathop{\rm Im}%
%EndExpansion
I)=\exp (\Delta S)\text{.}  \label{c23}
\end{equation}
This is a familiar result, it can be obtained by many other ways\cite{SM}%
\cite{PK}.

The integrand in (\ref{c18}) is quite complicate, it is not easy to
integrate directly, expanding$\ K(M)$ at the near field of $M_i$ as a Taylor
series
\begin{equation}
K(M)=K(M_i)+K^{^{\prime }}(M_i)(M-M_i)+\cdot \cdot \cdot \text{,}
\label{c24}
\end{equation}
we have
\begin{equation}
%TCIMACRO{\func{Im}}
%BeginExpansion
\mathop{\rm Im}%
%EndExpansion
I=-\left[ K(M_i)\Delta M+%
%TCIMACRO{\dfrac{K^{^{\prime }}(M_i)}2}
%BeginExpansion
{\displaystyle {K^{^{\prime }}(M_i) \over 2}}%
%EndExpansion
\Delta M^2+\cdot \cdot \cdot \right] =K(M_i)\omega -%
%TCIMACRO{\dfrac{K^{^{\prime }}(M_i)}2}
%BeginExpansion
{\displaystyle {K^{^{\prime }}(M_i) \over 2}}%
%EndExpansion
\omega ^2+\cdot \cdot \cdot \text{,}  \label{c25}
\end{equation}
where $\Delta M=M_f-M_i$\ is the variance of the mass of the black hole,
which should be negative, and $\omega =-\Delta M$ is the energy of the
radiated particle. Then the tunnelling\ rate
\begin{equation}
\Gamma \sim \exp [-2%
%TCIMACRO{\func{Im}}
%BeginExpansion
\mathop{\rm Im}%
%EndExpansion
I]=\exp \left\{ -2\left[ K(M_i)\omega -%
%TCIMACRO{\dfrac{K^{^{\prime }}(M_i)}2}
%BeginExpansion
{\displaystyle {K^{^{\prime }}(M_i) \over 2}}%
%EndExpansion
\omega ^2+\cdot \cdot \cdot \right] \right\} \text{.}  \label{c26}
\end{equation}
If we only take the first term while neglect the higher order terms in the
expression (\ref{c26}), that is
\begin{equation}
\Gamma \sim \exp \left[ -2K(M_i)\omega \right] .  \label{c27}
\end{equation}
We find that the the rate of tunnelling, as is expected, does take the form
of the Boltzmann factor $\exp (-\beta \omega )$ with the inverse of
temperature $\beta \equiv 1/T=2K(M)$. This is a familiar result and agrees
with Hawking's theory of the thermal radiation, so we confirm that Hawking
radiation do can be viewed as a process of particle tunnelling. But if we
take into account the higher order terms, thus the emission spectrum is not
precisely thermal, while yields corrections. This is a exciting news,
because whether the information is lost during the black hole radiation
depends in part on the fact that whether the spectrum is exactly thermal.
The spectrum is not thermal may open the way to looking for
information-carrying correlations in the spectrum.

\section{Radiation of charged particles}

\subsection{The velocity of a massive particle}

Since the trajectory followed by a massive particle is not light-like, it\
does not obey the equation of radial null geodesic, so equation (\ref{c10})
is no longer available. But for a massive particle, its 4-velocity has unit
length. In relativity, the tangent vector $U^a$ to a time-like curve
parameterized by the proper time $\tau $ is called the 4-velocity, it is
\begin{equation}
U^a=\left( \frac \partial {\partial \tau }\right) ^a.  \label{c28}
\end{equation}
The energy-momentum 4-vector, $p^a$, of the particle of ``rest mass'' $m$ is
defined by
\begin{equation}
p^a=mU^a,  \label{c29}
\end{equation}
and its covariant form(the dual vector)
\begin{equation}
p_a=g_{ab}p^b=mg_{ab}U^b.  \label{c30}
\end{equation}

For a massive particle moves in the space-time described by (\ref{c1}), the
``time component'' of $p_{^{_a}}$ is
\begin{equation}
p_0=mg_{0\nu }U^\nu =mg_{00}U^t=-E^{\prime }  \label{c31}
\end{equation}
that is
\begin{equation}
U^t\equiv \frac{dt}{d\tau }=\frac{-E^{\prime }}{mg_{00}}=\frac{-E}{g_{00}},
\label{c32}
\end{equation}
where $E$ is recognized to be the energy of per unit mass of the particle
measured by the stable observer at infinity.

According to the definition of the proper time $\tau $, the 4-velocity has
unit length, that is
\begin{equation}
U^aU_a=-1,  \label{c33}
\end{equation}
from (\ref{c1}), it can be written as
\begin{equation}
g_{00}\left( U^t\right) ^2+g_{11}\left( U^r\right) ^2=-1.  \label{c34}
\end{equation}
From (\ref{c32}) and (\ref{c34}) we have
\begin{equation}
U^r=\frac{dr}{d\tau }=\sqrt{g_{00}+E^2},  \label{c35}
\end{equation}
and
\begin{equation}
\frac{dr}{dt}=\frac{U^r}{U^t}=\frac{g_{00}\sqrt{g_{00}+E^2}}{-E}.
\label{c36}
\end{equation}
From (\ref{c8}) we get that
\begin{equation}
\frac{dT}{dt}=1+f(r)\frac{dr}{dt}=1+\frac{\sqrt{g}}{(1-g)\Delta }\frac{dr}{dt%
}.  \label{c37}
\end{equation}
Thus we get the radial velocity in coordinate $(T,r,\theta ,\varphi )$%
\begin{equation}
\dot{r}=\frac{dr}{dT}=\frac{dr}{dt}\frac{dt}{dT}=\frac{g_{00}\sqrt{g_{00}+E^2%
}}{-E-\sqrt{g}\sqrt{g_{00}+E^2}}=\frac{-\left( 1-\frac{r_{+}}r\right) \Delta
\sqrt{g_{00}+E^2}}{-E-\sqrt{\frac{r_{+}}r}\sqrt{g_{00}+E^2}}.  \label{c38}
\end{equation}

\subsection{Energy of GM space-time}

Obviously, the metric in (\ref{c1}) is static, and the time-like vector $\xi
^a=(\frac \partial {\partial t})^a$ is a killing vector, it's dual vector $%
\xi _a=g_{ab}\xi ^b=g_{00}(dt)_a$. For GM space-time, the determinant of the
metric is
\begin{equation}
g=-\left( r^2-D^2\right) ^2\sin ^2\theta \text{.}  \label{c39}
\end{equation}
The total energy(Komar energy\cite{Kom}\cite{Wald}) of a static,
asymptotically flat space-time which is vacuum in the exterior region is
defined as
\begin{equation}
E=-\frac 1{8\pi }\int_S\varepsilon _{abcd}\nabla ^c\xi ^d\text{,}
\label{c40}
\end{equation}
where $S$ is the boundary of a space-like hypersurface. The independence of
the right-hand side of equation (\ref{c40}) on the choice of $S$ depends on
the fact that $\xi ^a$ is a killing vector field. For GM space-time
\begin{eqnarray}
\nabla _a\xi _b &=&\nabla _a(g_{00}(dt)_b)=(\nabla
_ag_{00})(dt)_b+g_{00}(\nabla _a(dt)_b)  \nonumber \\
&=&(\nabla _ag_{00})(dt)_b+g_{00}(\partial _a(dt)_b-\Gamma _{ab}^c(dt)_c)
\nonumber \\
&=&(\frac d{dr}g_{00})(dr)_a(dt)_b-g_{00}\Gamma _{ab}^c(dt)_c\text{.}
\label{c41}
\end{eqnarray}
The second term
\begin{eqnarray}
g_{00}\Gamma _{ab}^c(dt)_c &=&g_{00}\Gamma _{\mu \nu }^\alpha \left( \frac %
\partial {\partial x^\alpha }\right) ^c\left( dx^\mu \right) _a\left( dx^\nu
\right) _b(dt)_c  \nonumber \\
&=&g_{00}\Gamma _{\mu \nu }^0\left( dx^\mu \right) _a\left( dx^\nu \right) _b%
\text{.}  \label{c42}
\end{eqnarray}
The nonvanishing components of the Christoffel symbol needed in (\ref{c42})
are
\begin{equation}
\Gamma _{01}^0=\Gamma _{10}^0=\frac 12g^{00}\frac d{dr}g_{00},  \label{c43}
\end{equation}
so that
\begin{eqnarray}
g_{00}\Gamma _{ab}^c(dt)_c &=&\frac 12g_{00}g^{00}\frac d{dr}g_{00}\left[
\left( dr\right) _a\left( dt\right) _b+\left( dt\right) _a\left( dr\right)
_b\right]  \nonumber \\
&=&\frac 12\frac d{dr}g_{00}\left[ \left( dr\right) _a\left( dt\right)
_b+\left( dt\right) _a\left( dr\right) _b\right] .  \label{c44}
\end{eqnarray}
That
\begin{eqnarray}
\nabla _a\xi _b &=&(\frac d{dr}g_{00})(dr)_a(dt)_b-\frac 12\frac d{dr}%
g_{00}\left[ \left( dr\right) _a\left( dt\right) _b+\left( dt\right)
_a\left( dr\right) _b\right]  \nonumber \\
&=&\frac 12(\frac d{dr}g_{00})\left[ \left( dr\right) _a\left( dt\right)
_b-\left( dt\right) _a\left( dr\right) _b\right]  \nonumber \\
&=&-\frac{\left[ MD^2+Mr^2-(P^2+Q^2)r\right] }{\left( r^2-D^2\right) ^2}%
\left[ \left( dr\right) _a\left( dt\right) _b-\left( dt\right) _a\left(
dr\right) _b\right] ,  \label{c45}
\end{eqnarray}
and
\begin{equation}
\nabla ^c\xi ^d=-\frac{\left[ MD^2+Mr^2-(P^2+Q^2)r\right] }{\left(
r^2-D^2\right) ^2}(\frac \partial {\partial r})^c\wedge (\frac \partial {%
\partial t})^d\text{.}  \label{c46}
\end{equation}
The volume element is
\begin{equation}
\varepsilon _{abcd}=\sqrt{-g}(dt)_a\wedge (dr)_b\wedge (d\theta )_c\wedge
(d\varphi )_d\text{.}  \label{c47}
\end{equation}
So from equation (\ref{c40}), we get that the total energy of Gibbons-Maeda
space-time is
\begin{eqnarray}
E &=&-\frac 1{8\pi }\int_S\frac{-\left[ MD^2+Mr^2-(P^2+Q^2)r\right] }{\left(
r^2-D^2\right) ^2}(\frac \partial {\partial r})^c\wedge (\frac \partial {%
\partial t})^d\times \sqrt{-g}(dt)_a\wedge (dr)_b\wedge (d\theta )_c\wedge
(d\varphi )_d  \nonumber \\
&=&-\frac 1{8\pi }\int_S\frac{-2\left[ MD^2+Mr^2-(P^2+Q^2)r\right] }{\left(
r^2-D^2\right) ^2}(d\theta )_a\wedge (d\varphi )_b=\frac{MD^2+Mr^2-(P^2+Q^2)r%
}{r^2-D^2}\text{.}  \label{c48}
\end{eqnarray}
From (\ref{c48}), we can see that the matter-gravity field and the
electromagnetic field have combined contribution to the total energy.

\subsection{Charged particle tunnelling across the horizon of GM black hole}

The GM black hole includes an axion dilaton charge and an electric charge,
so the study of the charged particle radiation is necessary. During the
tunnelling of the charged particle, the conservation of the total energy and
charge in the space-time plays an important role, that is to say ,when a
particle with static mass $\omega $, electro charge $q$ and magnetic charge $%
p$ tunnels across the event horizon, to the radiated particle, the
background metric was modified with $M,Q,P$ replaced by $M-\omega ,Q-q,P-p$.
For the radiation of charged particle, the action
\begin{eqnarray}
I &=&\int_{r_i}^{r_f}\int_{M_i}^{M_f}%
%TCIMACRO{\dfrac{dH}{\dot{r}}}
%BeginExpansion
{\displaystyle {dH \over \dot{r}}}%
%EndExpansion
dr=\int_{r_i}^{r_f}\int_{M_i}^{M_f}%
%TCIMACRO{\dfrac{dE(r)}{\dot{r}}}
%BeginExpansion
{\displaystyle {dE(r) \over \dot{r}}}%
%EndExpansion
dr  \nonumber \\
&=&\int_{r_i}^{r_f}\int_{M_i}^{M_f}%
%TCIMACRO{
%\dfrac{\left( -E-\sqrt{\frac{r_{+}}r}\sqrt{g_{00}+E^2}\right) r}{-\left( r-r_{+}\right) \Delta \sqrt{g_{00}+E^2}}}
%BeginExpansion
{\displaystyle {\left( -E-\sqrt{\frac{r_{+}}r}\sqrt{g_{00}+E^2}\right) r \over -\left( r-r_{+}\right) \Delta \sqrt{g_{00}+E^2}}}%
%EndExpansion
dE(r)dr=\int_{M_i}^{M_f}\int_{r_i}^{r_f}%
%TCIMACRO{\dfrac{f(r)dE(r)}{\left( r-r_{+}\right) }}
%BeginExpansion
{\displaystyle {f(r)dE(r) \over \left( r-r_{+}\right) }}%
%EndExpansion
dr,  \label{c49}
\end{eqnarray}
where
\begin{equation}
f(r)=%
%TCIMACRO{
%\dfrac{\left( -E-\sqrt{\frac{r_{+}}r}\sqrt{g_{00}+E^2}\right) r}{-\Delta \sqrt{g_{00}+E^2}}}
%BeginExpansion
{\displaystyle {\left( -E-\sqrt{\frac{r_{+}}r}\sqrt{g_{00}+E^2}\right) r \over -\Delta \sqrt{g_{00}+E^2}}}%
%EndExpansion
.  \label{c50}
\end{equation}
The integrand in (\ref{c49}) is singular at $r=r_{+}$, so just like equation
(\ref{c15})-(\ref{c17}), we get the imaginary part of the action
\begin{equation}
%TCIMACRO{\func{Im}}
%BeginExpansion
\mathop{\rm Im}%
%EndExpansion
I=-\pi \int_{M_i}^{M_f}f(r_{+})dE(r_{+}).  \label{c51}
\end{equation}

From (\ref{c48}), we get that
\begin{eqnarray}
dE(r) &=&\frac{\partial E}{\partial M}dM+\frac{\partial E}{\partial Q}dQ+%
\frac{\partial E}{\partial P}dP  \nonumber \\
&=&\left( dM-\frac{2Q}{r-D}dQ-\frac{2P}{r+D}dP\right)   \nonumber \\
&&-\frac{[MD^2+Mr^2-(P^2+Q^2)r]}{M(r^2-D^2)^2}\left(
2D^2dM+2DQdQ-2DPdP\right) ,  \label{c52}
\end{eqnarray}
and
\begin{eqnarray}
dE(r_{+}) &=&\left( dM-\frac{2Q}{r_{+}-D}dQ-\frac{2P}{r_{+}+D}dP\right)
\nonumber \\
&&-\frac{\sqrt{M^2+D^2-(P^2+Q^2)}}{M(r_{+}^2-D^2)}\left(
2D^2dM+2DQdQ-2DPdP\right) .  \label{c53}
\end{eqnarray}

From equation (\ref{c50}), we have
\begin{equation}
f(r_{+})=%
%TCIMACRO{
%\dfrac{\left( -E-\sqrt{E^2}\right) \left( r_{+}^2-D^2\right) }{-\sqrt{E^2}(r_{+}-r_{-})}}
%BeginExpansion
{\displaystyle {\left( -E-\sqrt{E^2}\right) \left( r_{+}^2-D^2\right)  \over -\sqrt{E^2}(r_{+}-r_{-})}}%
%EndExpansion
.  \label{c54}
\end{equation}
In equation (\ref{c32}), because $t$ and $\tau $ all increase towards the
future, so that $U^t>0$, while $g_{00}<0$, so we have $E>0$. The fact that
the energy of the particle $E>0$ is also reasonable in physics, because only
in the case $E>0$ the particle can have enough energy to move to infinity.
Then we have
\begin{equation}
f(r_{+})=%
%TCIMACRO{\dfrac{2(r_{+}^2-D^2)}{(r_{+}-r_{-})}}
%BeginExpansion
{\displaystyle {2(r_{+}^2-D^2) \over (r_{+}-r_{-})}}%
%EndExpansion
=%
%TCIMACRO{\dfrac{r_{+}^2-D^2}{\sqrt{M^2+D^2-(P^2+Q^2)}}}
%BeginExpansion
{\displaystyle {r_{+}^2-D^2 \over \sqrt{M^2+D^2-(P^2+Q^2)}}}%
%EndExpansion
.  \label{c55}
\end{equation}

From (\ref{c51}), (\ref{c53}), (\ref{c55}) we have
\begin{eqnarray}
%TCIMACRO{\func{Im}}
%BeginExpansion
\mathop{\rm Im}%
%EndExpansion
I &=&-\pi \int
%TCIMACRO{\dfrac{r_{+}^2-D^2}{\sqrt{M^2+D^2-(P^2+Q^2)}}}
%BeginExpansion
{\displaystyle {r_{+}^2-D^2 \over \sqrt{M^2+D^2-(P^2+Q^2)}}}%
%EndExpansion
\left( dM-\frac{2Q}{r_{+}-D}dQ-\frac{2P}{r_{+}+D}dP\right)   \nonumber \\
&&+\pi \int \frac 1M\left( 2D^2dM+2DQdQ-2DPdP\right) .  \label{c56}
\end{eqnarray}
In this equation, the second term of the right-hand-side
\begin{equation}
\pi \int \frac 1M\left( 2D^2dM+2DQdQ-2DPdP\right) =-\pi
\int_{(M_i,P_i,Q_i)}^{(M_f,P_f,Q_f)}d\left( D^2\right) =0,  \label{c57}
\end{equation}
here in the last step, we suppose that during the course of emission,
because of the interaction between the matter field and the electro-magnetic
field, the radiation of the mass, the electric charge and the magnetic
charge is not arbitrary, while obeys the regulation that keeping $D=\frac{%
P^2-Q^2}{2M}$ as a constant. Then we get
\begin{equation}
%TCIMACRO{\func{Im}}
%BeginExpansion
\mathop{\rm Im}%
%EndExpansion
I=-\pi \int
%TCIMACRO{\dfrac{r_{+}^2-D^2}{\sqrt{M^2+D^2-(P^2+Q^2)}}}
%BeginExpansion
{\displaystyle {r_{+}^2-D^2 \over \sqrt{M^2+D^2-(P^2+Q^2)}}}%
%EndExpansion
\left( dM-\frac{2Q}{r_{+}-D}dQ-\frac{2P}{r_{+}+D}dP\right) .  \label{c58}
\end{equation}

From (\ref{c20}), we have
\begin{equation}
\frac{\partial S}{\partial M}=%
%TCIMACRO{\dfrac{2\pi (r_{+}^2-D^2)}{\sqrt{M^2+D^2-(P^2+Q^2)}} }
%BeginExpansion
{\displaystyle {2\pi (r_{+}^2-D^2) \over \sqrt{M^2+D^2-(P^2+Q^2)}}}%
%EndExpansion
,  \label{c59}
\end{equation}
\begin{equation}
\frac{\partial S}{\partial Q}=-%
%TCIMACRO{\dfrac{2\pi (r_{+}+D)Q}{\sqrt{M^2+D^2-(P^2+Q^2)}} }
%BeginExpansion
{\displaystyle {2\pi (r_{+}+D)Q \over \sqrt{M^2+D^2-(P^2+Q^2)}}}%
%EndExpansion
,  \label{c60}
\end{equation}
\begin{equation}
\frac{\partial S}{\partial P}=-%
%TCIMACRO{\dfrac{2\pi (r_{+}-D)P}{\sqrt{M^2+D^2-(P^2+Q^2)}} }
%BeginExpansion
{\displaystyle {2\pi (r_{+}-D)P \over \sqrt{M^2+D^2-(P^2+Q^2)}}}%
%EndExpansion
.  \label{c61}
\end{equation}
So we have that
\begin{equation}
%TCIMACRO{\func{Im} }
%BeginExpansion
\mathop{\rm Im}%
%EndExpansion
I=-\frac 12\int \left( \frac{\partial S}{\partial M}dM+\frac{\partial S}{%
\partial Q}dQ+\frac{\partial S}{\partial P}dP\right) =-\frac 12%
\int_{S_i}^{S_f}dS=-\frac 12\Delta S.  \label{c62}
\end{equation}
Then for the radiation of charged particles, the tunnelling rate
\begin{equation}
\Gamma \sim \exp (-2%
%TCIMACRO{\func{Im} }
%BeginExpansion
\mathop{\rm Im}%
%EndExpansion
I)=\exp (\Delta S).  \label{c63}
\end{equation}
The result is just as same as that of the massless particle radiation.

\section{conclusion}

The calculations of the radiation of massless and charged particle from GM
black hole give the same conclusion that the self-gravitation of the
particle causes the radiation spectrum deviates from exact thermality and
the tunnelling rate satisfies $\Gamma \sim \exp (-2%
%TCIMACRO{\func{Im}}
%BeginExpansion
\mathop{\rm Im}%
%EndExpansion
I)=\exp (\Delta S)$. The later is a familiar result which was obtained by
many other methods. From quantum mechanics, the rate is expressed(Feynmn
Golden-law) as
\begin{equation}
\Gamma (i\rightarrow f)=\left| M_{fi}\right| ^2\cdot \text{(phase space
factor),}  \label{c64}
\end{equation}
where the first term on the right hand side is the square of the amplitude
for the process. The phase space factor is obtained by summing over final
states and averaging over initial states. Since the number of initial(final)
states is just the exponent of the initial entropy. So
\begin{equation}
\Gamma \sim \frac{e^{S_f}}{e^{S_i}}=e^{\triangle S}.  \label{c65}
\end{equation}
This implies that the radiation spectrum carries the microscopic information
of the black hole, the information is conservational.

\begin{acknowledgments}
This work is supported by the National Natural Science Foundation of
China under Grand No.10373003 and the National Basic Research
Program of China (No: 2003CB716300).
\end{acknowledgments}

\end{document}